\documentclass[finallayout,an,fleqn]{w-art}
\usepackage{times}
\usepackage{w-thm}
\theoremstyle{plain}

\theoremstyle{definition}

\usepackage[]{graphicx}
\chardef\bslash=`\\ 

\begin{document}
\DOIsuffix{theDOIsuffix}
\Volume{324}
\Issue{S1}
\Copyrightissue{S1}
\Month{01}
\Year{2003}
\pagespan{3}{}
\Receiveddate{15 November 2002}
\Reviseddate{30 November 2002}
\Accepteddate{2 December 2002}
\Dateposted{3 December 2002}



\title[Stellar Orbits]{Full Three Dimensional Orbits For Multiple Stars on 
Close Approaches to the Central Supermassive Black Hole}

\author[A. M. Ghez]{A. M. Ghez\footnote{Corresponding
     author: e-mail: {\sf ghez@astro.ucla.edu}, Phone: 310-206-0420,
     Fax: 310-206-2096}\inst{1}} \address[\inst{1}]{Department of Physics
and Astronomy, University of California, Los Angeles, CA 90095-162}
\author[E. Becklin]{E. Becklin\inst{1}}
\author[G. Duchene]{G. Duch\^ene\inst{1}}
\author[S. Hornstein]{S. Hornstein\inst{1}}
\author[M. Morris]{M. Morris\inst{1}}
\author[S. Salim]{S. Salim\inst{1}}
\author[A. Tanner]{A. Tanner\inst{1}}

\begin{abstract}
With the advent of adaptive optics on the W. M. Keck 10m telescope,
two significant steps forward have been taken in building the case for a 
supermassive black hole at the center of the Milky Way and understanding
the black hole's effect on its environment.  Using adaptive optics 
and speckle imaging to study the motions of stars in the plane of sky
with $\pm$$\sim$2 mas precision over the past 7 years, we have obtained 
the first simultaneous orbital solution for multiple stars. 
Among the included stars, three are newly identified (S0-16, S0-19, S0-20).
The most dramatic orbit is that of the newly identified star S0-16, which 
passed a mere 60 AU from the central dark
mass at a velocity of 9,000 km/s in 1999.
The orbital analysis results in a new central dark mass estimate of $3.6 
( \pm 0.4 ) \times 10^6 ((\frac{R_o}{8kpc})^3 M_{\odot}$.  This
dramatically strengthens
the case for a black hole at the center of our Galaxy, by confining
the dark matter to within a radius of
0.0003 pc or 1,000 R$_{sh}$ and thereby increasing the inferred dark mass 
density by
four orders of magnitude compared to earlier estimates.

With the introduction of an adaptive-optics-fed spectrometer, 
we have obtained the first 
detection of spectral absorption
lines in one of the high-velocity stars, S0-2, one month
after its closest approach to the Galaxy's central supermassive black hole.
Both Br $\gamma$ (2.1661 $\mu$m) and He I (2.1126 $\mu$m) are seen in
absorption with equivalent widths and
an inferred stellar rotational velocity that are consistent with that
of an O8-B0 dwarf, which suggests that S0-2 is
a massive ($\sim$15 $M_{\odot}$), young ($<$10 Myr) main sequence star.
Similarly, the lack of CO detected in our first AO spectra suggest
that several other of the high-velocity stars are also young.
This presents a major challenge to star formation theories, given the
strong tidal forces that prevail over all distances reached by these stars
in their current orbits and
the difficulty in migrating these stars inward during their lifetime
from further out where
tidal forces should no longer preclude star formation.

\end{abstract}

\maketitle                   

\section{Introduction}
While the Milky Way was neither the first nor the most obvious place to search
for a supermassive black hole, the case for one at the center of the Galaxy is
quickly becoming the most iron clad.  The first hint of a central concentration
of dark matter came from radial velocity measurements of ionized gas located in
a three-armed structure known as the mini-spiral, which extends from the center
out to about 1-2 pc (Lacy et al. 1980).   Concerns that the gases' motion are
not tracing
the gravitational potential were quickly allayed by radial velocity measurements
of stars, which are not susceptible to non-gravitational forces (McGinn et al.
1989; Haller et al. 1996; Genzel et al. 1997).
These early dynamical measurements of the gas and stars suggested the presence
of $3 \times 10^6 M_{\odot}$ of dark matter and confined it to within a radius
of $\sim$0.1 pc; the implied dark matter density was not sufficiently high to
definitively claim this as evidence for a single supermassive black hole,
since the measurements only imposed a lifetime for clusters of dark objects
of $5 \times 10^9 yrs$, which is not significantly shorter than the age of the
Galaxy (Maoz et al. 1998).    To make further progress in understanding the
underlying source of dark matter at the center of the Galaxy, it was necessary
to use techniques that compensated for the distorting effects of the Earth's
atmosphere, which had restricted the earlier studies of the dark matter
distribution to radii of 0.1 pc or larger.

In the early- to mid-1990's, two independent groups initiated 2 $\mu m$ high
spatial resolution imaging studies of the central stellar cluster to measure
the motions of stars in the plane of the sky.  While the ESO team began their
program using speckle imaging at the 3.6 m NTT and this year have moved to
the adaptive optics system on the 8 m VLT, the Keck team initiated their
program using speckle imaging on the 10 m Keck I telescope and began using
adaptive optics on the 10 m Keck II telescope in 1999.   The first phase of
these experiments yielded proper motion velocities, which increased the
implied dark matter density by 3 orders of magnitude to $10^{12} M_{\odot}/pc^3$
(Eckart \& Genzel 1997; Ghez et al. 1998).  This eliminated a cluster of dark
objects, such as neutron stars or stellar mass black holes, as a possible
explanation of the Galaxy's central dark mass concentration (Maoz et al.
1998) and left only
the fermion ball hypothesis (e.g., Tsiklauri \& Viollier 1998, Munyaneza \& 
Viollier 2002) as an alternative to a single supermassive black hole.   
The velocity dispersion measurements also localized the dark
matter to $\pm$100 mas (4 milli-pc) at a position consistent with the nominal
location of the unusual radio source Sgr A* (Ghez et al. 1998), whose
emission is posited to arise from accretion onto a central supermassive black
hole (e.g., Lo et al. 1985).   The proper motion experiments proceeded to strengthen both the
case for a supermassive black hole and its association with Sgr A* with the
detection of acceleration for three stars - S0-1, S0-2, and S0-4, which increased
the dark matter density to $10^{13} M_{\odot}/pc^3$ and positional accuracy
to $\pm$30 mas (Ghez et al. 2000; Eckart et al. 2002).    These experiments
also revealed that the orbital periods for S0-2 and S0-1 could be as short as
15 and 35 years, respectively, which would open a new arena for dynamical
studies of the central stellar cluster.

This paper summarizes the recent progress that has been made in this field
on two fronts with the W. M. Keck telescope.  
The first, with approximately a decade of proper motion measurements,
is the derivation of complete 3-dimensional orbits for multiple stars
that are making close approaches to the supermassive black
hole at the center of the Milky Way Galaxy.   The second is the measurement
of spectral lines in one of these high velocity stars.
These two steps forward make the strongest case yet for the
presence of a supermassive black hole at the center of the Galaxy
and, for the first time, allow us to take
an in-depth look at the question of where these stars formed.

\section{Observations \& Results}

\subsection{Proper Motions}

Beginning in 1995, K[2.2 $\mu m$]-band diffraction-limited images have been
obtained with the W. M. Keck I 10 m telescope to achieve an angular
resolution of 50 milli-arcsec and a positional accuracy of $\sim$2
milli-arcsec on stars located in the central
5$\tt'' \times$5$\tt''$
of our Galaxy.   
Since our original reporting of stars in this region (Ghez et al. 1998),
the sensitivity to high-velocity stars has been significantly improved.
Two primary factors contribute to
the increased number of recognized stars in this region.  First, the original
approach was conservative in an effort to avoid falsely identifying high
velocity stars on the basis of the three maps separated by a full year.  With
multiple observations each year beginning in 1998, it was clear that
many real sources were not being identified, leading to a decrease in the
threshold used to identify sources.  Second, in 1998 it became possible to
calibrate the alignment of the Keck telescope's mirror segments on NIRC,
the Keck I facility infrared camera,
allowing the telescope to be calibrated during the observing run at the
elevation of the Galactic Center, thereby significantly improving the image 
quality of our speckle maps.  
Figure 1 shows the three new proper motion stars (S0-16, S0-19, S0-20) and 
the one original proper motion star (S0-1, S0-2, S0-4) whose motions can now 
be modeled with Keplerian orbits.

Of particular note among the new sources (those with a label larger than 15)
is S0-19.  Its high proper
motion has caused it to be misidentified in earlier papers.  While it was
detected by us in 1995 as a K=14.0 source, two possible counterparts were
identified in 1996.  With limited time coverage, it was not possible to
definitively identify either as the correct counterpart and it was not
included in the proper motion sample (see discussion in Ghez et al. 1998).
The same source was reported as S3 (K=15) moving Westward by Eckart \& Genzel
(1997) and Genzel et al. (2000).
Genzel et al. (1997) report the detection of a new source, S12 (K$\sim$15), in
their 1996.43, located $\sim$0.$\tt''$1 of a source labeled S3 and proposed
it as the best candidate for the infrared counterpart of the compact radio
source Sgr A*.  This has been used in several recent papers to constrain
models of Sgr A*'s flared state (e.g., Narayan et al. 2002).  In our analysis,
it is now clear that S12 is simply a high velocity star that was coincident
with Sgr A* in 1996.  It is labeled S0-19 in this paper and it should be
associated with the 1992-1995 detection of S3.  The discrepancy in magnitudes 
arises from the difficulties of carrying out accurate photometry in such a 
crowded region.
This source illustrates the challenges associated with making a definitive
detection of infrared emission associated with Sgr A*, given the high stellar
densities and velocities and modest stellar intensity variations in this 
region.

\begin{figure}[htb]
\includegraphics[width=.45\textwidth]{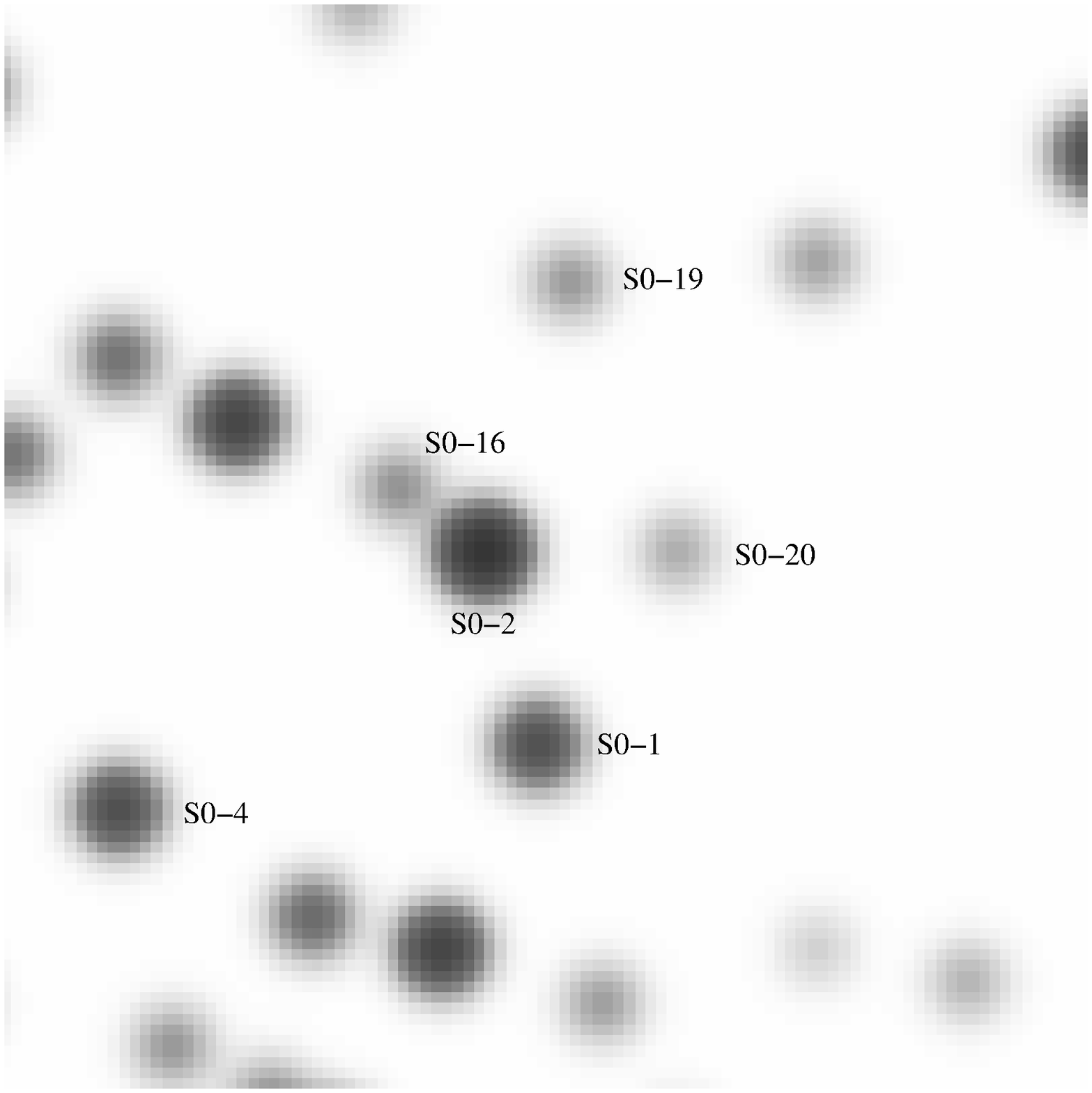}
\hfil
\includegraphics[width=.45\textwidth]{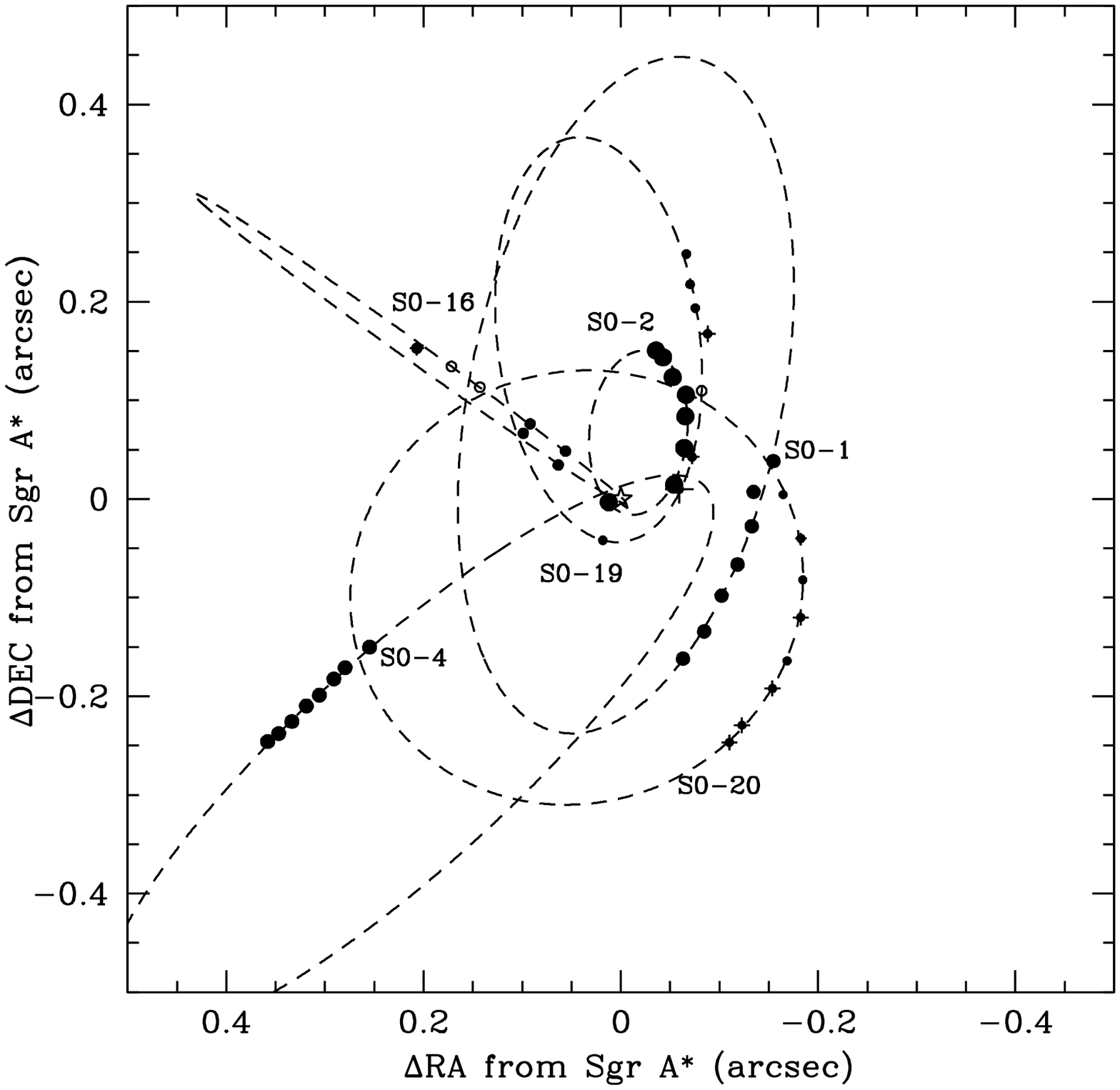}
\caption{(a) A 1$\tt" \times$1$\tt"$ cleaned image centered on the 
nominal position of Sgr A* showing the 2001 positions of some of the stars 
that have 
been followed over the course of the Keck proper motion study.   Three of the 
newly identified stars are S0-16, S0-19, and S0-20. (b) The annual positions
of some of these stars with orbital solutions.  Each star is labeled by its
first measurement. }
\label{fig:1}
\end{figure}

\subsection{Spectral Lines}

In 2002, the high proper motion star, S0-2, was observed with
NIRC2, the facility
near-infrared adaptive optics instrument (Matthews et al. 2003)
in a mode that achieved a spectral resolution of R$\sim$4000 ($\sim$75 km/s).
The resulting spectrum of S0-2, shown in Figure 2,
has two identifiable spectral lines.
These are both seen in absorption and are identified as
the H I (4-7) or Br$\gamma$ line at 2.166 $\mu$m and the He I
triplet at 2.1126 $\mu$m ($3p ~ ^3P^o-4s ~ ^3S$), which is a blend of
three transitions at 2.11274, 2.11267, and 2.11258 $\mu$m.
The detailed properties of these two lines, which are
obtained by fitting the background continuum over the whole
spectrum with a low-order polynomial and fitting the lines with a
Gaussian profile are reported in Ghez et al. (2003a).  

\begin{figure}[htb]
\includegraphics[width=\textwidth]{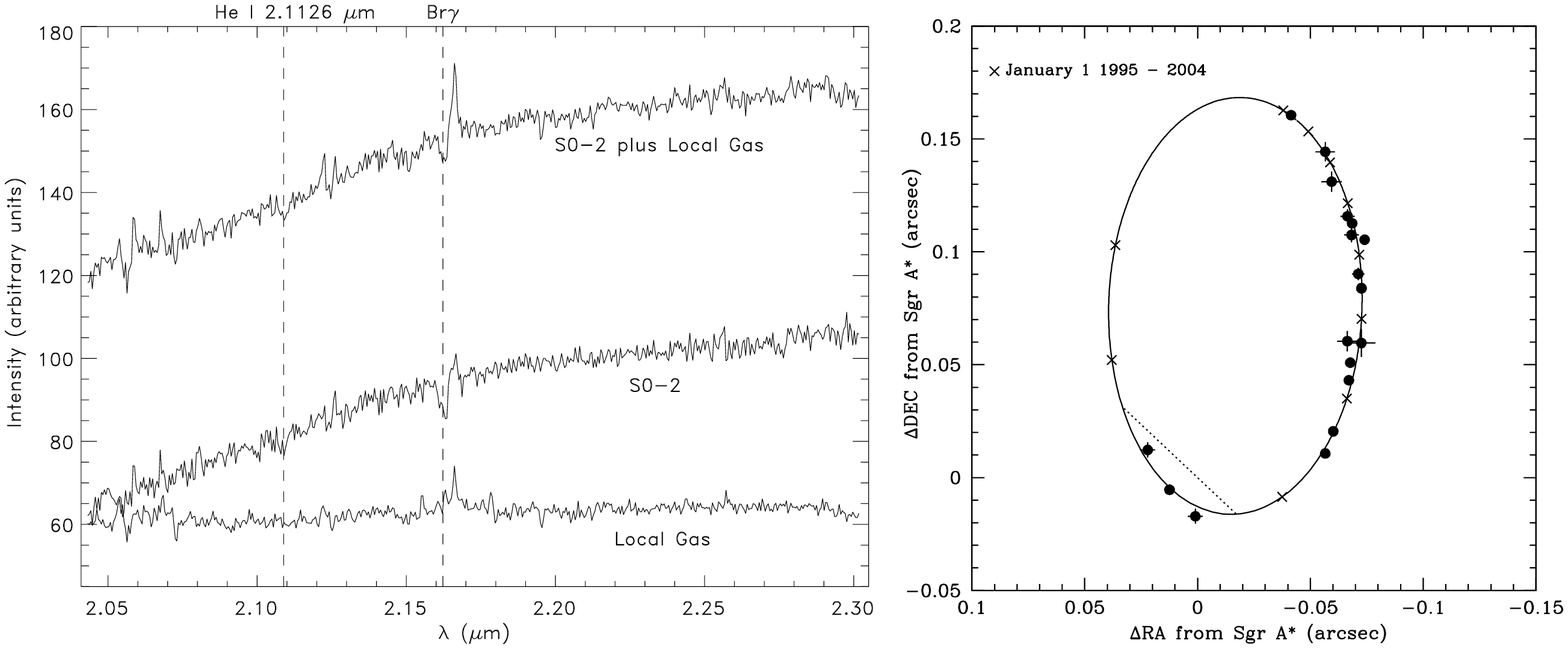}
\caption{
In the left panel is the first spectrum of S0-2 to show detectable
photospheric absorption lines (Br$\gamma$ and He I (2.1126~$\mu$m)).
The final spectrum (middle) is the raw
spectrum (top; with only an instrumental background removed)
minus a local sky (bottom).
The horizontal dimension has been re-binned by a factor of two for display
purposes only.
The vertical lines are drawn at 2.10899 and 2.16240 $\mu$m,
which correspond to the locations of Br$\gamma$ and He I for
a $V_{LSR}$ of -513 km/s.
This spectrum was obtained in 2000 June at the same time as one
of the proper motion measurements reported by Ghez et al. (2003)
and shown in the right panel (filled circles).  The crosses mark January 1 of
each year between 1995 and 2004 for the best fit orbit solution
(solid line), which
is based on both the radial velocity and proper motions.  The dotted
line is the line of nodes, which reveals S0-2 to be behind the black hole
for a mere $\sim$0.5 years out of its 15 year orbit. Adapted from Ghez et al.
(2003)
}
\label{fig:2}
\end{figure}

\section{Discussion \& Conclusions}

\subsection{Dynamics}

The strong deviations from linear motions on the plane of the sky 
for stars within 0.$\tt"$2 of Sgr A* along with 
the radial velocity measurements for S0-2 provide new and powerful constraints
on their orbital parameters, which are presented in Ghez et al. (2003a, b) and
Ghez (2003).
In the orbital fits, we assume that (1) the stellar masses are insignificant
compared to a central point source, (2) the central point source has no
significant velocity with respect to the Galaxy, which is supported by the lack 
of motion detected for Sgr A* by Reid et al. (1999) and Reid (2003),
and (3) the central point
source has a distance of 8.0 kpc (Reid 1993).
This leaves the following 9 unknowns:  Center of Attraction ($x_o$, $y_o$), 
Period (P), Semi-Major Axis
(A), Eccentricity (e), Time of periapse passage (T$_o$), Angle of nodes
to periapse ($\omega$), Angle of the line of nodes ($\Omega$), and Inclination (
i).  We begin by fitting each star independently.
Each of these stars has recently gone through
periapse (1999 - 2002).  In the most extreme case, S0-16 passed within
a mere 60 AU of the central dark mass, while traveling at a velocity
of 9,000 km/sec.  The solution for S0-2 is
consistent with that reported by Sch\"odel et al. (2002)
and, despite the two additional free parameters introduced by fitting
for the center of attraction, the uncertainties on the orbital
parameters are reduced by a factor of 2-3. 
Since the independent centers of attraction are consistent with one another,
we proceed to fit the orbital motion for all the stars simultaneously with a
common center of attraction.  This first orbital estimate of the Galaxy's
dynamical center is not only consistent with the nominal
infrared position of Sgr A* to within the uncertainties on the latter
(Reid et al. 2003), but is also a factor of 7 more precise 
($\pm$ 1.5 milli-arcsec).   Furthermore, the agreement between the 
masses inferred from the simultaneous Keplerian orbit fits for multiple 
stars (see Figure 3 and Table 1) suggests that the central dark mass
potential is well modeled by a point source with mass
$3.6 ( \pm 0.4) \times 10^6 (D/8kpc)^3
M_{\odot}$, consistent at the $\sim$2$\sigma$ level with
earlier estimates based on velocity dispersion measurements.
These measurement increase the dark mass density by four orders of magnitude,
ruling out Fermion balls as an alternative hypothesis for all 
supermassive black holes (Viollier 2003).

\begin{figure}[htb]
\begin{minipage}{0.65\textwidth}
\includegraphics[width=\textwidth]{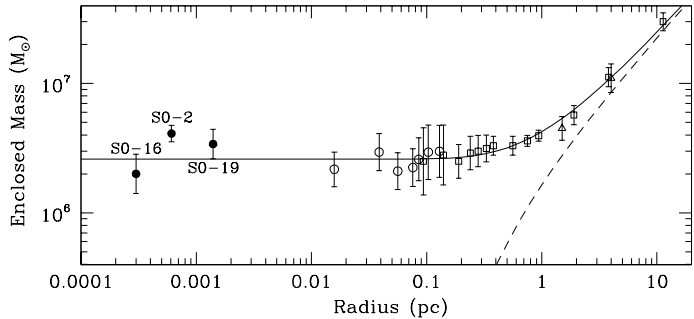}
\caption{ Enclosed mass as a function of radius.  
The masses from the individual star's orbital motion agree both with one 
another and the earlier estimates based on velocity dispersion measurements. 
The solid line shows the best fit black hole plus luminous cluster model 
based on the earlier measurements.  The new orbital 
masses increases the central dark mass density by 4 orders of magnitude, 
dramatically
strengthening the case for a central supermassive black hole.}
\label{fig:1}
\end{minipage}
\hfil
\begin{minipage}{0.25\textwidth}
\setfloattype{table}
\caption{Estimates of the central dark mass from fits to the stellar orbital 
motion.  The reported values come from fits that solve for a common center 
of attraction and assume a distance of 8 kpc. 
Only solutions with fractional 
uncertainties less than 30\% are listed here.
}
\label{tab:1}\renewcommand{\arraystretch}{1.5}
\begin{tabular}{ll} \hline
Star & Mass ($10^6 M_{\odot}$)\\ \hline
S0-2 & 4.1 $\pm$ 0.6 \\
S0-16 & 3.0 $\pm$ 0.7 \\
S0-19 & 3.4 $\pm$ 0.9 \\
Average   & 3.6 $\pm$ 0.4 \\
\end{tabular}
\end{minipage} 
\end{figure}

For S0-2, the addition of radial velocity measurements also breaks the
ambiguity in the inclination angle, {\it i}.
With the proper motion data alone, only the
absolute value of the inclination angle can be determined,
leaving the questions of the direction of revolution and
whether the star is located behind the black hole at periapse
(closest approach) unresolved.
Our radial velocity measurements indicate a negative inclination angle
and consequently that S0-2 is both counter-revolving against the Galaxy
and behind the black hole at the time of periapse.
The improved location of the center of attraction 
from the orbital analysis
results in a minimum offset of S0-2 from the black hole
in the plane of the sky of 14 $\pm$ 2 milli-arsec, which is
significantly larger than the expected Einstein radius
(R$_E$ = 0.42 milli-arcsec for an assumed distance behind the black hole
of $\sim$ 100 AU) and therefore makes
gravitational lensing a negligible effect (Wardle \& Yusef-Zadeh 1992;
Alexander \& Loeb 2001).

In principle, the addition of radial velocities to the study of S0-2's dynamics
allows the distance to the Galactic Center, $R_o$, to be a free parameter in
the orbital fits (Salim \& Gould 1999).  The measurements, however, were
obtained just 30 days after the star's
closest approach to the black hole when the radial velocity was changing very
rapidly (see Figure 4).
While the current radial velocity and proper motion data set constrains
$M/{R_o}^3$ very effectively ($\sim$15\% uncertainty), it does not yet
produce a meaningful measurement
of $R_o$.  Nonetheless, as Figure 4 shows, the radial velocities from the
currently allowed orbits quickly diverge, producing a spread
of a few hundred km/s in one year.
Within the next few years,  the orbital fits based on both proper motions
and additional radial velocity measurements should provide the most
direct and precise estimate of the distance to the Galactic Center, making
it a fundamental rung in the cosmic distance ladder.

\begin{vchfigure}[htb]
\includegraphics[width=0.6\textwidth]{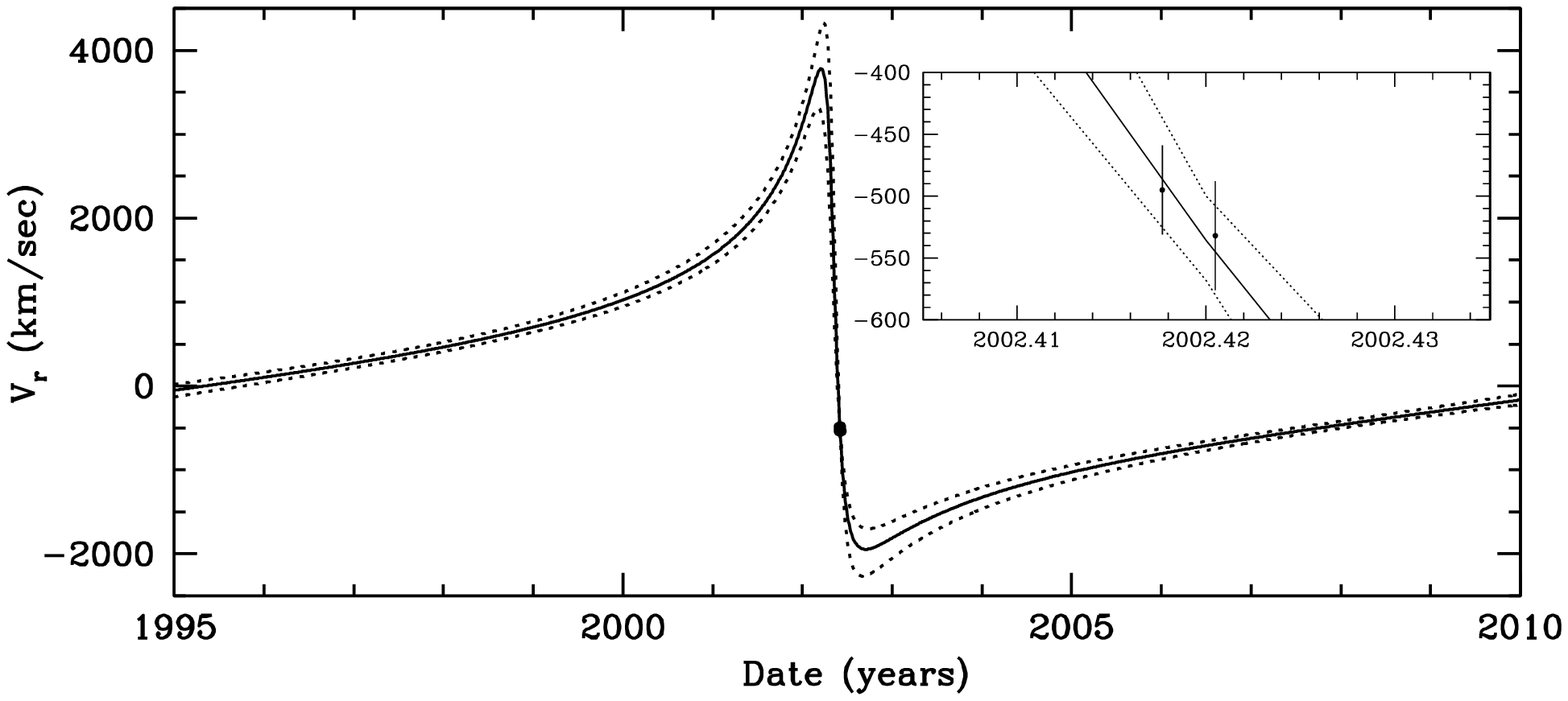}
\vchcaption{
The measured radial velocity along with the predicted
radial velocities.   The solid curve comes from the best fit orbit and the
dotted curves display the range for the orbital solutions
allowed with the present data sets. Adapted from Ghez et al. (2003a)
}
\label{fig:4}
\end{vchfigure}

\subsection{Stellar Astrophysics}

The detection of absorption lines in S0-2 allows us to sort out the spectral
classification ambiguities present when only photometric information
is available and to determine if this star's photosphere has been
altered as a result of its close proximity to the central black hole
(Ghez et al. 2003a).
The average brightness at 2.2 $\mu$m for S0-2 is K $\sim$ 13.9
mag and there is no evidence of brightening after periapse passage
(Ghez et al. 2003b).  With a distance of 8.0 kpc and
K-band extinction of 3.3 mag (Rieke, Rieke, \& Paul 1989),
the 2.2 $\mu$m brightness of S0-2 implies that, if it is an ordinary star
unaltered by its environment, it could either be an O9 main-sequence star
or a K5 giant star;
all supergiants are ruled out as they are too bright by at least 2 magnitudes
in the K bandpass.
Kleinmann and Hall (1986) provide a 2.0 - 2.5 $\mu$m spectral atlas of
late-type stars that demonstrates that if S0-2 is a K5 giant star, then it
should have deep CO absorption lines, which definitively were not detected
in either this experiment or our earlier
experiment reported by Gezari et al. (2002).  In contrast,
the spectral atlas of 180 O and B stars constructed by Hanson, Conti and
Rieke (1996) shows that an O9 main sequence star both lacks the
CO absorption and has Br $\gamma$ and He I (2.1126 $\mu$m) consistent
with the observed values.
Furthermore, stars earlier than O8 
in this comparison sample
show NIII (2.115 $\mu$m) in emission and
He II (2.1885 $\mu$m) in absorption above our 3 $\sigma$ thresholds;  
the lack of photospheric He I (2.058 $\mu$m) absorption
does not provide any additional constraints.
Similarly, dwarf B-type stars later than B0
have absorption-line  equivalent widths that are too large.
Together, the photometry
and absorption line-equivalent widths permit dwarf spectral types ranging
from O8 to B0.  Likewise, the rotational velocity of 224 km/s is reasonable for
this range (Gatheier, Lamers, \& Snow 1981).
S0-2, therefore,
appears to have a spectral type, and hence effective temperature ($\sim$30,000 K),
as well as luminosity ($\sim$10$^3$ $L_{\odot}$) that are consistent
 with a main sequence star having a mass of
$\sim$15 $M_{\odot}$ and an age $<$10 Myr.

It is challenging to explain the presence of such a young star
in close proximity to a supermassive black hole.
Assuming that the black hole has not significantly affected
S0-2's appearance or evolution, S0-2 must be younger than
10 Myr and thus formed relatively recently.   If it has not
experienced significant orbital evolution, its apoapse distance
of 1900 AU implies that star formation is possible in spite of
the tremendous tidal forces presented by the black hole, which
is highly unlikely.  If the star formed at larger distances from the
black hole and migrated inward, then the migration would have to be through a
very efficient process.  Current understanding of the distribution of stars,
however, does not permit such efficient migration.
This problem is similar to that raised by the He I emission-line stars
(e.g., Sanders 1992,1998;
Morris 1993, Morris et al. 1999; Gerhard 2001; Kim \& Morris 2002),
which are also counter-revolving against the Galaxy (Genzel et al. 1997),
but amplifies it with a distance from the black hole that is an order of
magnitude smaller.
An alternative explanation for S0-2's hot photosphere is that it
 may be significantly altered by its environment.
While its periapse passage is too large for it to be tidally heated
by the black hole as explored by Alexander \& Morris (2003), it may be
affected by the high stellar densities found in this region.
On the one hand, the high stellar densities might allow S0-2 to be
an older giant star that has had
its outer atmosphere stripped through collisions; however, to generate
the necessary luminosity,  significant external heating is required
(Alexander 1999).  On the other hand, high stellar densities might
lead an unlikely capture of a component in a massive binary star system
(Gould \& Quillen 2003) 
or a cascade of merger events (Lee 1996), which which would allow S0-2's formation process to have begun more than
10 Myr ago.
However a large number of collisions would have had to occur to
provide the necessary lifetime to bring it in from sufficiently
large radii.
More exotically, it could be a "reborn"
star, which occurs as the product of a merger of a stellar remnant with
a normal star.   None of these possibilities is altogether satisfactory,
leaving the Sgr A* cluster stars as a paradox of apparent youth in the
vicinity of a supermassive black hole.

\begin{acknowledgement}
This work has been supported by the National Science Foundation through
the individual grant AST99-88397 and the Science and Technology Center for
Adaptive Optics, managed by the University of California at Santa Cruz under
Cooperative Agreement No. AST - 9876783.
The W.M. Keck Observatory is operated as a scientific partnership among the
California Institute of Technology, the University of California and the
National Aeronautics and Space Administration. The Observatory was made
possible by the generous financial support of the W.M. Keck Foundation.
\end{acknowledgement}


\begin{thebibliography}{10}

\bibitem{bib1} Alexander, T. 1999, ApJ, 527, 835

\bibitem{} Alexander, T., \& Loeb, A. 2001, ApJ, 551, 223

\bibitem{bib2} Alexander, T., \& Morris, M. 2002,
in prep

\bibitem{bib4} Eckart, A., \& Genzel, R. 1997, MNRAS, 284, 576

\bibitem{bib5} Eckart, A., Genzel, R., Ott, T., \&
Sch\"odel, R. 2002, MNRAS, 331, 917


\bibitem{bib8} Genzel, R., Eckart, A., Ott, T., \& Eisenhauer, F. 1997, MNRAS, 291, 219

\bibitem{} Genzel, R., Pichon, C., Eckart, A., Gerhard, O. E., Ott, T.,
2000, MRAS, 317, 348


\bibitem{bib9} Gerhard, O. 2001, ApJ, 546, L39

\bibitem{bib10} Gezari, S., Ghez, A.~M., Becklin, E.~E.,
Larkin, J., McLean, I.~S., Morris, M. 2002, ApJ, 576, 790

\bibitem{} Ghez, A.~M. 2003, Carnegie Observatories Astrophysics Series, Vol. 1: Coevolution of Black Holes and Galaxies, ed. L. C. Ho 
(Cambridge: Cambridge Univ. Press) 

\bibitem{} Ghez, A.~M. et al. 2003a, ApJLett, in press  
(astro-ph/030229).

\bibitem{bib11} Ghez, A.~M., Hornstein, S., Salim, S., Tanner, A., Morris, M.,
and Becklin, E.~E. 2003b, in prep

\bibitem{bib12} Ghez, A.~M., Klein, B.~C., Morris, M.,
\& Becklin, E.~E. 1998, ApJ, 509, 678

\bibitem{bib13} Ghez, A.~M., Morris, M., Becklin,
E.~E., Tanner, A., \& Kremenek, T. 2000, Nature, 407, 349

\bibitem{} Gould, A., \& Quillen, A. 2003, ApJ, submitted (astro-ph/0302437)

\bibitem{bib15} Hanson, M.~M., Conti, P.~S.,
 \& Rieke, M.~J. 1996, ApJS, 107, 281

\bibitem{bib16} Kim, S.~S., \& Morris, M.
2002, ApJ, in press

\bibitem{bib17} Kleinmann, S.~G., \& Hall, D.~N.~B
1986, ApJS, 62, 501

\bibitem{bib18} Lee, H.~M., 1996, IAU 169, 215

\bibitem{bib19} Lo, K.~Y., Backer, D.~C.,
Ekers, R.~D., Kellermann, K.~I., Reid, M., \& Moran, J.~M.\ 1985, Nature,
315, 124


\bibitem{} Maoz, E. 1998, ApJ, 494, 181L

\bibitem{bib20} Matthews, K. et al. 2003, PASP, in prep

\bibitem{bib21} Morris, M., 1993, ApJ, 408, 496

\bibitem{bib22} Morris, M., Ghez, A.~M., Becklin, E.~E.
1999, Adv. Spa. Res., 23, 959

\bibitem{} Munyaneza, F., Viollier, R. D. 2002, ApJ, 564, 274.

\bibitem{bib23} Reid, M.~J. 1993, ARA\&A, 31, 345

\bibitem{} Ried, M.~J. 2003, this proceedings

\bibitem{bib24} Reid, M.~J., Readhead, A.~C.~S., Vermeulen,
R.~C., Treuhaft, R.~N. 1999, ApJ, 524, 816

\bibitem{bib25} Reid, M.~J., Menten, K.~M., Genzel, R., Ott, T.,
Sch\"odel, R., \& Eckart, A. 2003, ApJ, submitted

\bibitem{bib26} Rieke, G.~H., Rieke, M.~J., \& Paul, A.~E.
1989, ApJ, 336, 752

\bibitem{bib28} Salim, S., \& Gould, A. 1999, ApJ,
523, 633

\bibitem{bib29} Sanders, R.~H. 1992, Nature, 359, 131

\bibitem{bib30} Sanders, R.~H. 1998, MNRAS, 294, 35

\bibitem{bib31} Sch\"odel, R. et al.  2002, Nature, 419, 694

\bibitem{} Tsiklauri, D., Viollier, R.~D. 1998, ApJ, 500, 591

\bibitem{bib33} Wardle, M. \& Yusef-Zadeh, F. 1992, ApJ, 387, L65.

\end{thebibliography}
\end{document}